\documentstyle[12pt,aaspp4,epsfig]{article}

\begin{document}

\def \cmm  {cm$^{-2}$ }
\def \cmmm {cm$^{-3}$ }
\def \kms  {km~s$^{-1}$ }
\def \Lya  {Ly$\alpha$ }
\def \Lyb  {Ly$\beta$ }
\def \Lyg  {Ly$\gamma$ }

\def \object {1157+3143 }

\title{The Detection of Two Distinct High Ionization States in a 
	QSO Lyman Limit Absorption System:  Evidence for Hierarchical 
	Galaxy Formation at $z \sim 3$?}

\author{David Kirkman\altaffilmark{1,2}
\& David Tytler\altaffilmark{1,3}\\
Department of Physics, and Center for Astrophysics and Space Sciences;
\\ University of California, San Diego; \\ MS 0424; La Jolla; CA
92093-0424\\}

\altaffiltext{1} {Visiting Astronomer, W.M. Keck Observatory which
is a joint facility of the University of California, the California
Institute of Technology and NASA.}

\altaffiltext{2} {E-mail: dkirkman@ucsd.edu}
\altaffiltext{3} {E-mail: tytler@ucsd.edu}

\abstract 

We have detected two high ionization phases of gas in the $z \sim
2.77$ partial Lyman limit system (LLS) towards QSO \object.  We detect
the first phase by CIV and SiIV absorption, and the second phase --
which is either warmer or undergoing larger random bulk motions than
the first -- via OVI absorption.  Both phases of gas are present in
similar column density ratios in each of the 5 velocity components,
making it appear that this LLS is constructed of 5 very similar
building blocks.  We find that this system displays some of the
properties expected of a hierarchical merging event, although
published models may have trouble explaining the SiIV absorption we
observe.  When different ions show similar velocity structure, we
commonly assume that they arise in the same gas, and we compare their
column densities to derive the ionization and abundances.  For this
one absorption system the different ions have similar velocity but they
do not arise in the same gas.


\section {Introduction}

Absorption systems with enough neutral Hydrogen to be optically thick
to Lyman continuum radiation are commonly found in the spectra of high
redshift QSOs.  These Lyman limit (LLS) and damped Lyman alpha (DLA)
systems often show line absorption from multiple ionization states of
several different elements, most commonly Carbon and Silicon.  It is
often assumed that optically thick absorption systems contain gas
in two separate phases -- gas that is shielded from UV background Lyman
continuum radiation and thus contains low ions such as CII and SiII,
and gas that is not completely shielded from the UV background and thus
shows high ionization species such as CIV and SiIV.  The low ions are
believed to be in a separate phase from the high ions because in DLA
systems all low ions have a similar velocity structure that appears to
have no relation to the velocity structure of the high ions, all of
which also have a similar velocity structure (Prochaska and Wolfe 1997).

In recent years, many groups have begun to use numerical simulations
to simulate the formation of structure in variants of cold dark matter
(CDM) Universes (Cen et al. 1994; Zhang, Anninos, \& Norman 1995;
Hernquist et al. 1996).  These simulations naturally produce absorbers
that at least superficially resemble LLS and DLA systems as a natural
result of the hierarchical structure formation that takes place in CDM
models (Katz et al. 1996; Gardner et al. 1997).  In most simulations
LLS and DLAs have very similar physical structures and tend to be
protogalaxies in the process of formation via the merging of several
smaller structures.

There is not complete agreement within the community that the
structures identified in the CDM simulations accurately represent the
absorbers seen in actual QSO spectra.  In particular, Prochaska and
Wolfe (1997) find that the kinematics of DLA low ionization gas is
consistent with the gas being located in a thick rotating disk -- a
picture that is quite different from the simulation results.
Nonetheless, the advent of numerical simulations has sparked a
renaissance in the study of QSO absorption systems, and the
simulations are now making detailed predictions about the observed
properties of absorption systems, including the metal lines in
DLAs and LLS.

Rauch et al. (1997) calculated the expected metal absorption from DLA
and LLS systems found in a CDM simulation with the assumption that the
absorbing gas had a uniform metallicity.  They found that while the low
ions are all found in the same gas, the gas producing most of the SiIV
absorption is not the same gas producing C IV or O VI absorption.  As
expected, they found that absorption from low ions occurs in only the
highest density gas near the center of the protogalactic structures.
The high ions were found in lower density gas surrounding the
protogalactic clump, in a structure resembling an onion.  SiIV was
found only very close to the center, much like the low ions.  CIV was
found primarily in low density gas further out, and OVI was found
mainly in even lower density gas further yet from the center of the
clump.  SiIV and CIV were found in gas inside the shock front of a
collapsing object, whereas OVI was found in gas still falling into
the structure.  This resulted in OVI lines that were systematically
wider than the CIV or SiIV lines.

In this letter we present the first multiple component OVI absorption
features to be detected in a high $z$ intervening absorption system.
The only other OVI detection at high $z$ (Kirkman and Tytler, 1997)
contained only one component, and only the $\lambda$ 1032 line of the
$\lambda$ 1032, 1039 doublet was detected because of blending.  Here
we report the detection of at least 5 OVI components in both members
of the $\lambda$ 1032, 1039 doublet in the $z \sim 2.77$ LLS towards
\object. We also see SiIV, CIV, CII, SiII, and SiIII.  This system is
remarkable because, as we will show, all high ions have a very similar
velocity structure, yet we are able to show on observational grounds
alone that the OVI absorption cannot all be occurring in the same gas
producing the CIV and SiIV absorption.

\section {The $z \sim 2.77$ LLS}

We observed QSO \object for a total of 7 hours in March 1996 and
January 1997 with the HIRES spectrograph (Vogt 1992) on the W.M. Keck
Telescope during our search for deuterium.  The resulting spectrum has
a resolution of 7.9 \kms and a SNR $\sim 50$ per 0.03 \AA~ pixel.  The
spectrum was reduced and extracted in the standard fashion
(e.g. Kirkman and Tytler, 1997).

There are two LLS in this spectrum, one at $z \sim 2.94$ and another
at $z \sim 2.77$. The $z \sim 2.94$ LLS in this spectrum shows neither
OVI nor CIV and will not be discussed.  We do not have a good
estimate of the HI column for the $z \sim 2.77$ LLS because the $z
\sim 2.94$ LLS blots out the spectrum below 3600 \AA~ which prevents
us from seeing any Lyman series line higher than \Lyg.  Nonetheless,
we believe the $z\sim2.77$ system is a Lyman limit because it shows
both CII and SiII, neither of which are expected to be present in an
absorption system that is not optically thick to Lyman continuum
radiation.  The lines associated with the $z \sim 2.77$ partial LLS 
found in this spectrum are shown in Figure 1.  

Figure 2 shows the OVI and CIV lines of this system in more detail.
We used VPFIT (Webb, 1987) to fit Voigt profiles to the lines shown in
Figure 2; the line parameters appear above each line in the figure.
While fitting, we did not in any way tie line parameters between
different elements -- the OVI fit is completely independent of the
CIV fit.  Although there is not a 1-to-1 correspondence between the
OVI and CIV lines in Figure 2, it is clear that the velocity
structure of the OVI lines does trace the velocity structure of the
CIV lines.  Figure 2 also gives the impression that the OVI
absorption is a smeared out version of the CIV absorption.  This
impression is confirmed by noting that the $b$ values are larger for
OVI than for CIV for the absorbers centered near $-140$ \kms and 40
\kms, which are well defined and lightly blended lines in all three
species.

There is more than apparent similarity between the OVI and CIV lines
in this absorption system.  In Figure 3, we show that the OVI lines
can be accurately fit using only profiles restricted to be at the same
velocity as the identified CIV lines.  While the velocities of the OVI
lines in Figure 3 were fixed, the column densities and $b$ values of
each line were allowed to vary to produce the best match to the observed
spectrum.  Note that in the final fit the $b$ value in each OVI component
is $\sim 2$ times as large as in the corresponding CIV component, and
that the column density ratio N(OVI)/N(CIV) varies only by $\sim 5$
between the different components.

\section{Ionization State of the $z \sim 2.77$ LLS }

The ionization state of this system is not well constrained by the
available data.  This is because the OVI and CIV line widths in this
system prevent us from using the commonly made assumption that all of
the lines at the same velocity arise in the same gas.  Thus we can not
use the column density ratios of the observed ions to work out the
ionization state of the system using assuming either photo or
collisional ionization equilibrium.  As demonstrated in Figure 3, the
OVI line is wider that the CIV line in each velocity component of
this LLS.  Since Oxygen is heavier than Carbon, this means that some
or all of the OVI absorption is arising in gas different than that
producing the CIV absorption -- there are at least two phases of gas
in this system.  The phase which produces the OVI absorption is
either warmer or more turbulent than the phase producing the CIV
absorption.

All of the OVI lines associated with this system have $b > 17$ \kms,
which is equivalent to T $> 2.8 \times 10^5$ K if the line widths are
thermal. An ionization parameter of $U > 1$ is required for the UV
background (assuming the spectrum is a QSO like $-1.5$ power law) to
photoheat gas to this temperature (Donahue and Shull, 1991).  Since it
is unlikely that gas associated with a LLS is this rarefied, the large
OVI $b$ values probably mean that the lines are widened by non-thermal
motions, are collisionally ionized, or are actually several lines so
close to each other in velocity space that their individual profiles
can not be resolved.

Except for the components at $-50$ and $+65$ \kms, all of the CIV
lines have $6 < b < 8$ \kms.  This corresponds to a maximum
temperature of $2.6 - 4.6 \times 10^4$ K.  This is too cold for
collisionally ionized gas to show CIV absorption, so the CIV lines
must be photoionized.  It is tempting to use the observed column
densities of CIV, SiIV, SiII and CII to constrain the ionization
states of the velocity components of this system by assuming they are
in photoionization equilibrium with the UV background.  However, we
feel that this would be a mistake.  The main result we draw from this
data is that each velocity component of the LLS contains at least two
phases of gas.  We feel that this result warrants re-examination of
the commonly made assumption that all of the absorption seen at the
same velocity arises in the same gas.  In particular, if the OVI and
CIV absorption do not arise in the same gas, there is no good reason
to assume that the CIV and SiIV absorption comes from the same gas,
and good reason to suspect that the CIV and CII absorption does not
arise in the same gas.

\section{Discussion}

There are several scenarios that can give rise to multiple high
ionization phases of gas in an individual component of an absorption
line system.  The absorbing gas may contain pockets of cool or warm
(CIV, SiIV) gas intermixed with pockets of hot gas (OVI).  In this
scenario a method must be found to widen the OVI lines; presumably
they would arise in small pockets of shock heated gas and/or expanding
gas around collapsing substructure within each component (stars?).
The similar metal column ratios imply the structure within each of
the components is similar as well.

In a more likely scenario, the multiple phases are explained by
density gradients in the absorbing components at each velocity.
Moving along the gradient, the photoionization parameter of the gas
will change, giving a large number of effective gas phases.  As
discussed in the introduction, this sort of situation was produced in
the simulations run by Rauch et al.  In this scenario, the wide OVI
lines are produced primarily because the OVI absorbing gas is falling
into the collapsing structure, whereas the CIV absorbing gas is at
rest behind the shock front.  The observed wide OVI lines and the
similar component structure of this LLS suggest we may be observing
the formation of a protogalaxy by the merging of several smaller
structures.  There is, however, at least one major difference between
the data and the Rauch et al. simulation.  In this system, the SiIV
absorption is strong and has the same velocity structure as the other
high ions.  In the simulations this will only occur if each velocity
component is centered on the line of sight to the QSO because of the
small effective cross section of SiIV absorbing gas from the center of
a collapsed object.  This seems improbable.  It will be interesting to
see if the OVI lines observed at high resolution in other LLS agree
with the physical picture of the LLS suggested by these observations.

\acknowledgements

We thank Tom Bida and Barbara Schafer with the W.M. Keck Observatory
for assistance with our observations of \object, and Tom Barlow for
providing a copy of his extraction software which allowed us to
rapidly do science with our data.  We thank Bob Carswell for making
his VPFIT software package available to us.  This work was supported
in part by NSF grant 31217A and by NAGW4497 from NASA.

\clearpage


\begin{figure}
\centerline{\psfig{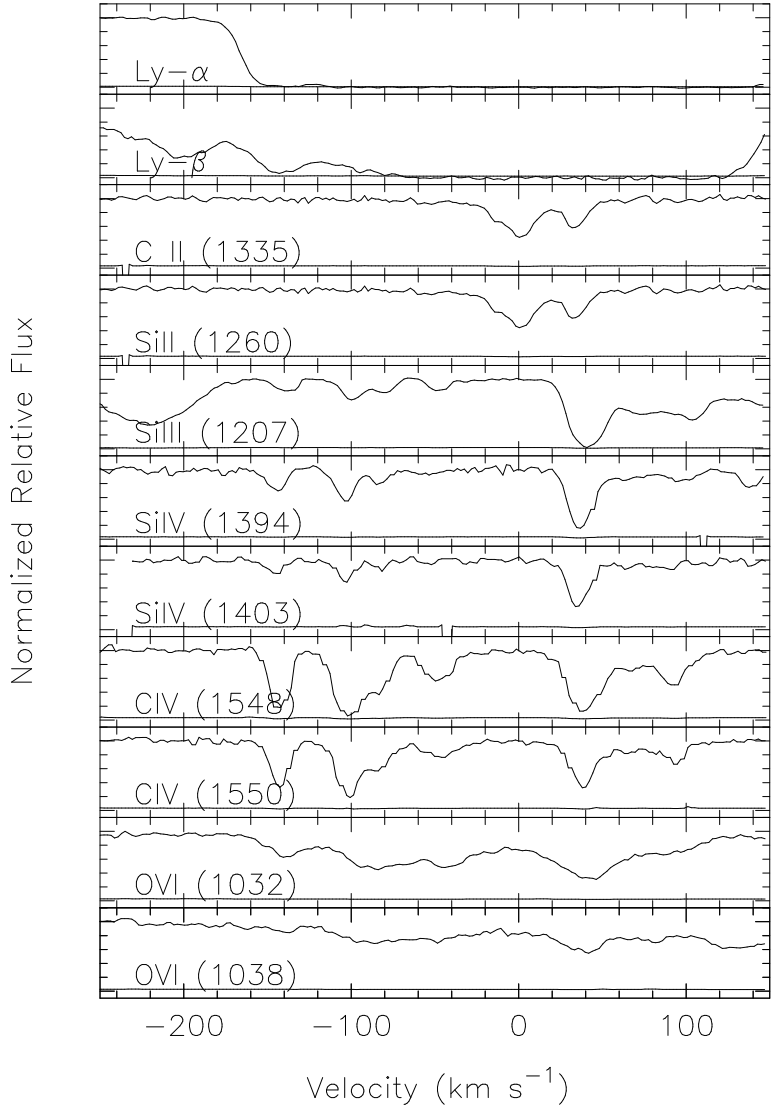}}
\caption{\label{oviIIfigI} All identified metal lines associated 
        with the $z \sim 2.77$ LLS.}
\end{figure}

\begin{figure}
\centerline{\psfig{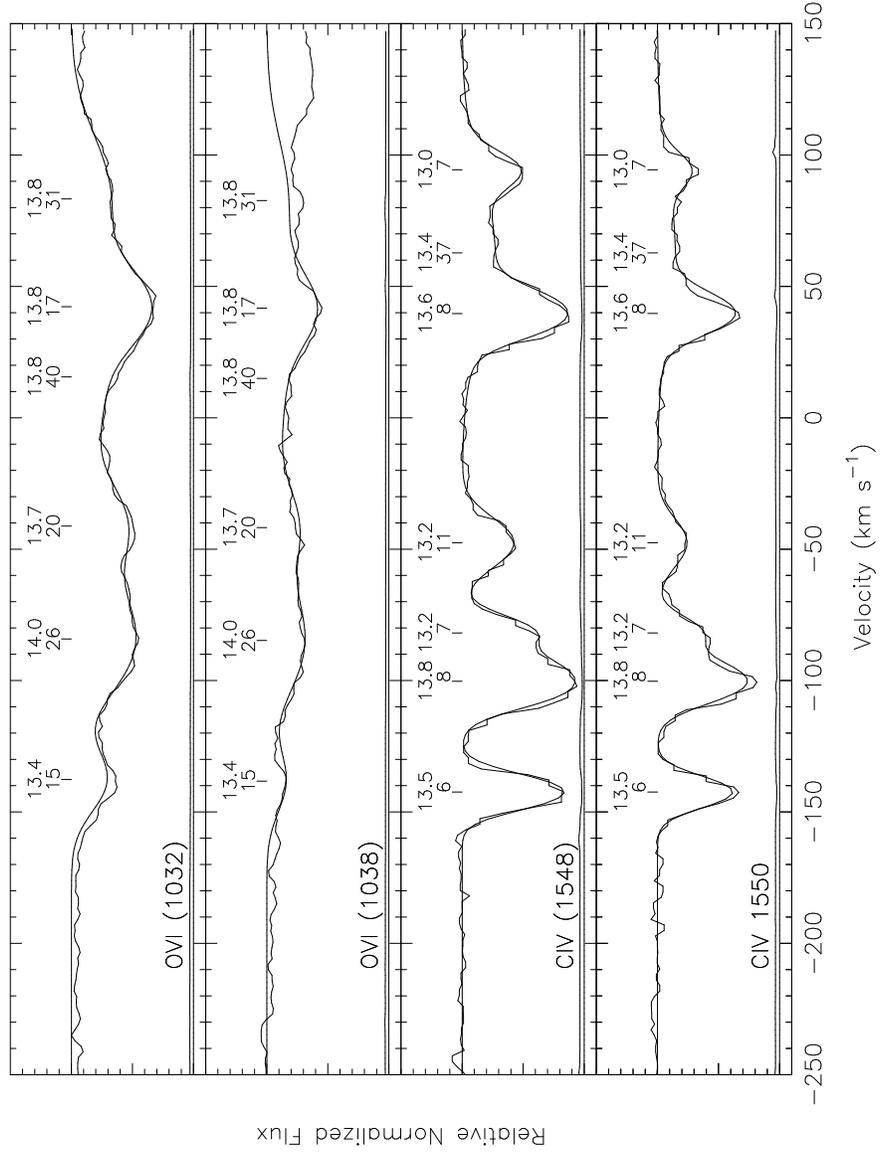}}
\caption[Independent Voigt profile fits to the OVI and CIV
        lines associated with the $z \sim 2.77$
        LLS.]{\label{oviIIfigII} Independent Voigt profile fits to the
        OVI and CIV lines associated with the $z \sim 2.77$ LLS.  The
        column density and velocity dispersion are displayed on top of
        each line.}
\end{figure}

\begin{figure}
\centerline{\psfig{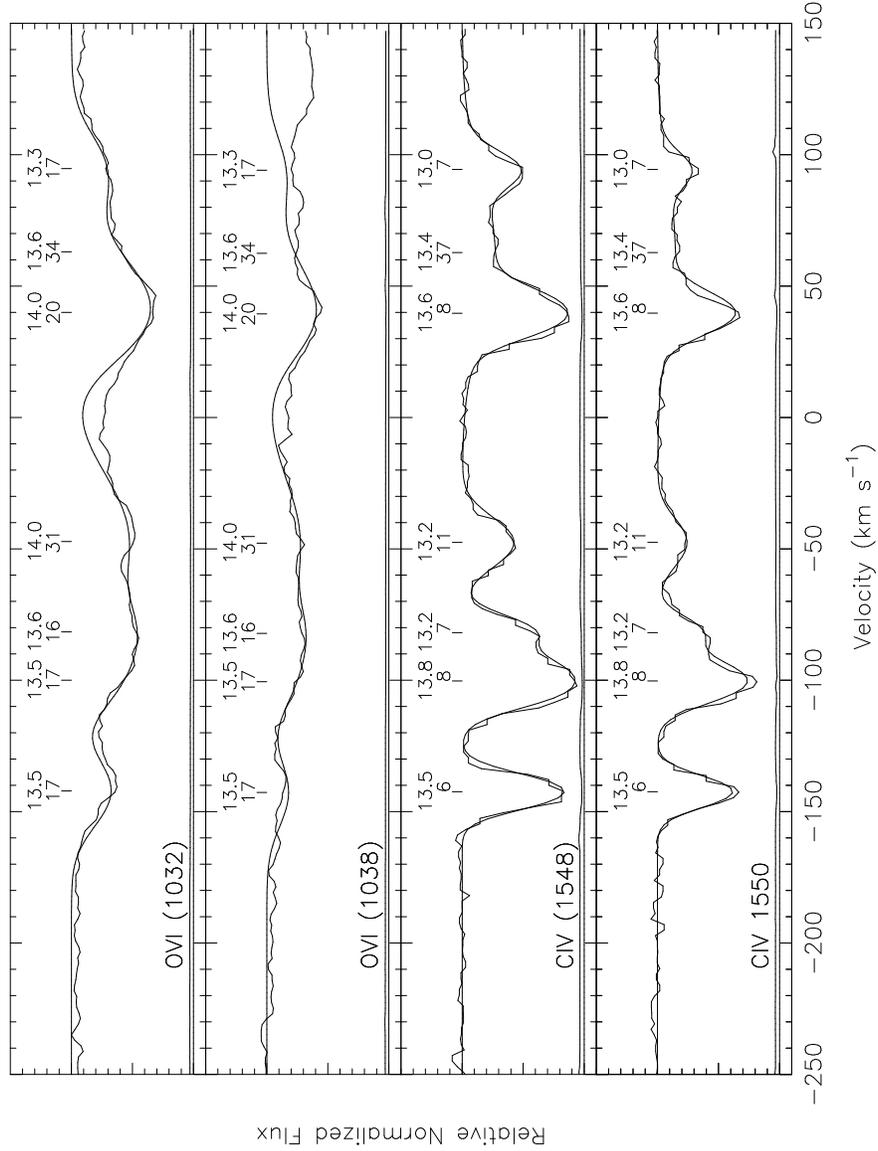}}
\caption[Voigt profile fits to the OVI and CIV absorption with
        OVI components only at the position of CIV components
        identified in Figure 5.2.]{\label{oviIIfigIII} Voigt profile
        fits to the OVI and CIV absorption with OVI components only at
        the position of CIV components identified in Figure 5.2.  With
        the exception of a small region near $v \sim 0$ \kms, all of
        the OVI absorption can be accurately fit.  To produce an
        accurate fit, however, requires that $b_{OVI} > 2 b_{CIV}$ in
        each component.  We therefore conclude that while the OVI
        absorption closely traces the CIV absorption in velocity
        space, the OVI lines are systematically wider than the CIV
        lines.}
\end{figure}




\end{document}